\pdfoutput=1
\documentclass[aps, pra,
superscriptaddress,
amsmath,amssymb,
longbibliography,
reprint
]{revtex4-2}
\usepackage{graphicx}

\usepackage{textcomp, gensymb}

\usepackage{bm}
\usepackage{dcolumn}
\usepackage[utf8]{inputenc}
\usepackage[T1]{fontenc}
\usepackage{mathptmx}
\usepackage{etoolbox}
\usepackage{hyperref}
\hypersetup{
    colorlinks=true,
    allcolors=blue
    }

\begin{document}


\title{Shallow Silicon Vacancy Centers with lifetime-limited optical linewidths in Diamond Nanostructures}
\author{Josh A. Zuber}
\thanks{These authors contributed equally.} 
\affiliation{Department of Physics, University of Basel, CH-4056 Basel, Switzerland}
\affiliation{Swiss Nanoscience Institute, University of Basel, CH-4056 Basel, Switzerland}
\author{Minghao Li}
\thanks{These authors contributed equally.} 
\affiliation{Department of Physics, University of Basel, CH-4056 Basel, Switzerland}
\author{Marcel.li Grimau Puigibert}
\affiliation{Department of Physics, University of Basel, CH-4056 Basel, Switzerland}
\author{Jodok Happacher}
\affiliation{Department of Physics, University of Basel, CH-4056 Basel, Switzerland}
\author{Patrick Reiser}
\affiliation{Department of Physics, University of Basel, CH-4056 Basel, Switzerland}
\author{Brendan J. Shields}
\affiliation{Department of Physics, University of Basel, CH-4056 Basel, Switzerland}
\author{Patrick Maletinsky}
\email{patrick.maletinsky@unibas.ch}
\affiliation{Department of Physics, University of Basel, CH-4056 Basel, Switzerland}
\affiliation{Swiss Nanoscience Institute, University of Basel, CH-4056 Basel, Switzerland}

\date{July 24, 2023}


\begin{abstract}
The negatively charged silicon vacancy center (SiV$^-$) in diamond is a promising, yet underexplored candidate for single-spin quantum sensing at sub-kelvin temperatures and tesla-range magnetic fields.
A key ingredient for such applications is the ability to perform all-optical, coherent addressing of the electronic spin of near-surface SiV$^-$ centers.
We present a robust and scalable approach for creating individual, $\sim$50\,nm deep SiV$^-$ with lifetime-limited optical linewidths in diamond nanopillars through an easy-to-realize and persistent optical charge-stabilization scheme. The latter is based on single, prolonged 445\,nm laser illumination that enables continuous photoluminescence excitation spectroscopy, without the need for any further charge stabilization or repumping. Our results constitute a key step towards the use of near-surface, optically coherent SiV$^-$ for sensing under extreme conditions, and offer a powerful approach for stabilizing the charge-environment of diamond color centers for quantum technology applications.
\end{abstract}

\maketitle


Diamond color centers represent the backbone for many research directions in quantum technologies, including sensing\,\cite{hedrichNanoscaleMechanicsAntiferromagnetic2021,bianNanoscaleElectricfieldImaging2021,neumannHighPrecisionNanoscaleTemperature2013,vindoletOpticalPropertiesSiV2022,liuSiliconVacancyNanodiamondsHigh2022}, quantum information processing\,\cite{hegdeEfficientQuantumGates2020}, and quantum communication\,\cite{bradleyTenQubitSolidStateSpin2019,englundDeterministicCouplingSingle2010,pompiliRealizationMultinodeQuantum2021}. 
In quantum sensing, the optically addressable electron spin of the nitrogen vacancy (NV) center has been successfully employed to sense various physical observables, including magnetic fields\,\cite{hedrichNanoscaleMechanicsAntiferromagnetic2021}, electric fields\,\cite{bianNanoscaleElectricfieldImaging2021}, and temperature\,\cite{neumannHighPrecisionNanoscaleTemperature2013}. In particular, scanning probe magnetometry based on single NV centers offers quantitative imaging with nanoscale resolution that enabled insights into physical systems that are inaccessible to classical approaches\,\cite{hedrichNanoscaleMechanicsAntiferromagnetic2021,thielProbingMagnetism2D2019}. 
However, the deployment of NV magnetometry in extreme conditions, such as mK temperatures and tesla-range magnetic fields is hampered by the near-surface NV's charge instability in cryogenic environments and limitations in coherent driving its electronic spin, which requires driving fields of tens of GHz in frequency. 
Yet, nanoscale sensing under such conditions would offer exciting opportunities to address interesting condensed matter systems, such as fractional quantum Hall effects\,\cite{bolotinObservationFractionalQuantum2009}, or unconventional superconductors\,\cite{ishidaSpintripletSuperconductivitySr2RuO41998}, by direct, nanoscale magnetic imaging.

The negatively charged silicon vacancy center (SiV$^-$) is an alternative diamond color center hosting an electronic spin that offers promising and advantageous properties for single-spin quantum sensing under extreme conditions. 
Compared to the NV center, the SiV$^-$ predominantly emits photons in the zero phonon line (ZPL) and thereby presents a more efficient spin-photon interface\,\cite{nguyenQuantumNetworkNodes2019}.
Moreover, the SiV$^-$ orbital and spin level structure generally allows for all-optical coherent driving of its ground-state spin\,\cite{beckerAllOpticalControlSiliconVacancy2018,beckerUltrafastAllopticalCoherent2016,pingaultAllOpticalFormationCoherent2014}.

\begin{figure}[!hbt]
    \centering
    \includegraphics[width=82.55mm]{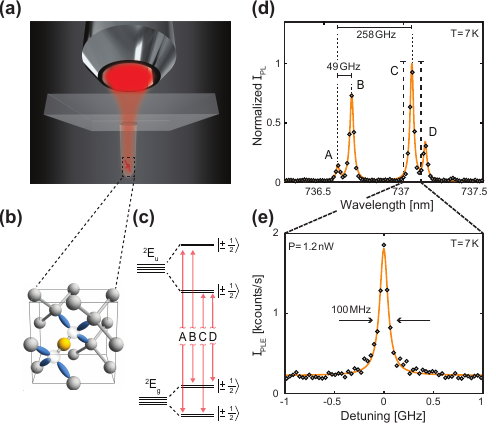}
    \caption{Optically coherent SiV$^-$ in nanostructured diamond. 
(a) Rendering of the sample geometry, with an emitter placed at the focus of an overhanging parabolic reflector (PR), and optical addressing performed through the diamond substrate. 
(b) Atomic structure of SiV$^-$, with the color center symmetry axis being oriented along the diamond 111-axis. 
The Si atom (yellow) is located at the interstitial site between two C vacancies (transparent). Nearest-neighbour C dangling bonds shown in blue. 
(c) Zero-field energy level diagram of SiV$^-$. 
Red arrows denote the four zero-field optical transitions labelled A, B, C and D. 
(d) Typical optical spectrum of a single SiV$^-$ center obtained with off-resonant laser excitation (wavelength $\lambda=515$\,nm) at 7\,K, showing the zero-field optical transitions as well as the ground and excited state splittings of $\sim$ 49\,GHz and $\sim$ 258\,GHz, respectively. 
Data in black dots and a four-peak Lorentzian fit in yellow. 
(e) Charge repump-free photoluminescence excitation (crf-PLE) measurement with 1.2\,nW resonant laser power sweeping across transition C while recording the phonon sideband (PSB) intensity.
Data were acquired by 14 successive laser frequency sweeps over six minutes.
The Lorentzian fit (yellow) to the data (black dots) reveals a lifetime-limited linewidth of $100.4\pm 6.9$\,MHz.}
 \label{Fig:1Overview}
\end{figure}

The inversion symmetry of SiV$^-$ leads to a vanishing electric dipole moment and renders its optical transition frequency insensitive to electric field fluctuations to first order\,\cite{rogersMultipleIntrinsicallyIdentical2014}. As a result, highly coherent photon emission with linewidths limited by the inverse excited state lifetime ($\sim1.7$\,ns) have been observed for SiV$^-$ centers far from the diamond surface\,\cite{schroderScalableFocusedIon2017}, or even in diamond nanocrystals\,\cite{hausslerPreparingSingleSiV2019}.
Together with the SiV$^-$ center's substantial electronic spin coherence times at mK temperatures\,\cite{sukachevSiliconVacancySpinQubit2017}, these properties open the exciting perspective to perform all-optical, coherent, nanoscale quantum sensing with SiV$^-$ spins.
Realizing this potential requires the ability to create SiV$^-$ centers with high optical coherence within a few tens of nanometers from the diamond surface in nanostructures suited for efficient sensing operation\,\cite{hedrichParabolicDiamondScanning2020}.
However, this achievement has remained elusive so far, largely 
because shallow SiV$^-$ suffer from significant spectral instability induced by electric field noise originating from 
nearby diamond surfaces\,\cite{evansNarrowLinewidthHomogeneousOptical2016a,machielseQuantumInterferenceElectromechanically2019}, which is further exacerbated by diamond nanofabrication. 
Furthermore, resonant excitation of SiV$^-$, essential for all-optical sensing schemes, usually requires off-resonant charge-resetting laser pulses\,\cite{evansNarrowLinewidthHomogeneousOptical2016a,nicolasSubGHzLinewidthEnsembles2019,langLongOpticalCoherence2020,arjonamartinezPhotonicIndistinguishabilityTinVacancy2022}, that lead to additional spectral diffusion\,\cite{chuCoherentOpticalTransitions2014} and laser heating, both of which form further obstacles to the use of SiV$^-$ for quantum sensing. Here, we present a reproducible approach to address these challenges and to realize shallow ($\lesssim$ 50\,nm deep), single SiV$^-$ centers with high optical coherence in diamond nanopillars shaped into parabolic reflectors (PRs)\,\cite{hedrichParabolicDiamondScanning2020}. Our approach to SiV$^-$ creation and diamond nanofabrication produces a $\sim30\%$ yield in creating close to lifetime limited SiV$^-$ centers.\\

Importantly, we additionally introduce an easy-to-realize charge stabilization procedure that enables such narrow linewidths in close to $100\%$ of the shallow SiV$^-$ centers in our PRs. This charge stabilization consists of a single, prolonged exposure of SiV$^-$ to 445\,nm laser light, which has the striking effect of narrowing the transition linewidths for SiV$^-$ exhibiting initially broad lines. 445\,nm laser illumination furthermore enables continuous photoluminescence excitation (PLE) measurements without any need for optical charge repumping -- a PLE scheme that we refer to as charge-repump-free PLE (crf-PLE) and which we discuss further below. Most SiV$^-$ we investigated after this procedure show inhomogeneous linewdiths that fall within an approximate factor of two of the lifetime limit, with several instances showing near-lifetime limited single sweep linewidths. In Fig.\,\ref{Fig:1Overview}(e), we present the narrowest transition linewidth achieved with our approach, which shows a full width at half maximum (FWHM) Lorentzian linewidth of $\Delta\nu=100.4\pm6.9$\,MHz.

\begin{figure}[!ht]
    \centering
    \includegraphics[width=82.55mm]{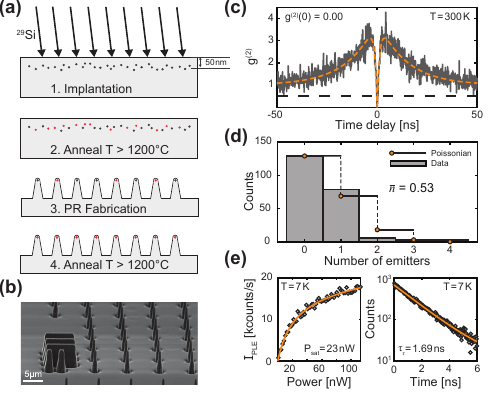}
    \caption{Sample fabrication and optical properties of SiV$^-$ in parabolic reflectors (PR). 
(a) Sample preparation workflow. 
First, we implant the diamond with $^{29}$Si$^+$ at a dose of $6 \times 10^9$ ions/cm$^2$ at an angle of 7\,\degree and an energy of 80\,keV. 
Then we anneal the diamond in a home-built high vacuum oven to produce SiV$^-$. 
Third, we nanofabricate parabolic reflectors and subsequently repeat the annealing procedure from step 2 to further increase the yield of SiV$^-$ formation and to enhance optical coherence\,\cite{evansNarrowLinewidthHomogeneousOptical2016a}. Grey dots are Si ions while red dots indicate successfully formed SiV$^-$.
(b) SEM image of a PR array (imaging angle 70\degree). 
Binary bulk markers on the sample are visible to the left. 
(c) Exemplary room-temperature (RT) background corrected off-resonant $g^{(2)}(\tau)$ recorded on a PR by exciting the emitter with a 515\,nm diode laser and recording ZPL photoluminescence (PL) intensity. 
The fit (dashed yellow) to the data (grey) reveals $g^{(2)}(0)=0.00\pm0.16$, indicating a single emitter. 
(d) SiV$^-$ number distribution and a corresponding Poissonian fit with mean $\bar{n}=0.53$ emitters per pillar. 
(e) Left panel: low-temperature (LT) resonant saturation curve on transition C recorded by tuning the diode laser into resonance and varying its optical power using an AOM while collecting PSB photons. 
Saturation power for this particular SiV$^-$ is $23.0\pm3.1$\,nW. 
Right panel: LT ZPL PL decay of the SiV$^-$ recorded with off-resonant pulsed excitation at 515\,nm. 
Fitting (yellow line) reveals a typical optical lifetime of $1.69\pm0.04$\,ns.}
 \label{Fig:2Fab}
\end{figure}


\section*{\label{sec:SamplePrep}Results}

\textbf{Shallow and coherent SiV$^-$ in parabolic reflectors}

 Our diamond preparation and nanofabrication procedure is outlined in Fig.\,\ref{Fig:2Fab}(a): We begin with a commercially available electronic grade diamond (Element Six), sample A, which we implant (CuttingEdge Ions) with $^{29}$Si$^+$ ions at an angle of 7\degree, a dose of $6 \times 10^9$ ions/cm$^2$, and an implantation energy of 80\,keV. This energy yields a Stopping Range of Ions in Matter (SRIM) predicted emitter depth of $\sim$50\,nm [supplementary information (SI) Fig.\,S1]. In order to form SiV$^-$, we anneal the implanted diamond in a home-built high-vacuum oven at 400\,\degree C, 800\,\degree C and 1300\,\degree C for 4\,h, 11\,h and 2\,h respectively.
This corresponds to a slight modification of the procedure introduced by Evans et al.\,\cite{evansNarrowLinewidthHomogeneousOptical2016a}, where we increase the temperature of the last annealing step, as it has been shown that higher temperatures benefit the optical coherence of 
SiV$^-$\,\cite{langLongOpticalCoherence2020}. 
Successful SiV$^-$ creation is confirmed by observing its room-temperature (RT) ZPL around 738\,nm, under off-resonant optical excitation at a wavelength of $\lambda=515$\,nm.
Subsequently, in order to enhance the emitters' collection and excitation efficiencies, we nanofabricate parabolic reflectors (PRs) with diameters at the apex of $\sim 300$\,nm on the sample. 
For this, we use electron-beam lithography defined SiO$_x$ etch masks and a sequence of plasma etching steps (detailed elsewhere\,\cite{hedrichParabolicDiamondScanning2020}). 
PRs are arranged in arrays [Fig.\,\ref{Fig:2Fab}(b)] to facilitate both the fabrication procedure and systematic characterization of SiV$^-$. 
The fabricated PRs employ the same design otherwise used for diamond scanning tips in scanning NV magnetometry\,\cite{appelFabricationAllDiamond2016}, which will expedite future use of SiV$^-$ for scanning probe microscopy.

After diamond nanofabrication, we perform a second anneal identical to the first one, as Evans et al.\,\cite{evansNarrowLinewidthHomogeneousOptical2016a} have shown that the optical coherence of SiV$^-$ increases by removing sub-surface damage from the diamond lattice by annealing and subsequent acid cleaning. Additionally, we observe a significant increase of SiV$^-$ yield after the second annealing step, as many PRs do not show a SiV$^-$ ZPL immediately after fabrication at the implantation dose we employed. Thus the annealing steps before and after PR fabrication are a crucial ingredient for creating individual and coherent SiV$^-$ centers in nanostructures.\\

To characterize our PR arrays, we perform systematic measurements at RT using an automized, home-built confocal microscopy setup (see Methods). 
We measure optical spectra, off-resonant saturation curves [SI Fig.\,S2] and off-resonant second order correlation functions $g^{(2)}(\tau)$ for each pillar in the array. An exemplary data set of a background corrected $g^{(2)}(\tau)$ recorded from a PR is shown in Fig.\,\ref{Fig:2Fab}(c) with a fit revealing $g^{(2)}(0)=0.00\pm0.16$, indicating the presence of a single emitter in the PR. 
For background correction, we subtract from the raw autocorrelation data the uncorrelated background signal stemming from background photons, whose intensity we determined by recording photoluminescence saturation curves [see SI Fig.\,S2] -- a procedure proposed earlier by Brouri et al.\,\cite{brouriPhotonAntibunchingFluorescence2000} [SI Sec.~IV]. 
Subsequently, we estimate the number of emitters in a PR using the relationship $g^{(2)}(0)=1-\frac{1}{n}$, where $n$ is the number of emitters\,\cite{brouriPhotonAntibunchingFluorescence2000}, while in the absence of a SiV$^-$ ZPL, we assign $n=0$ to the PR. 
Using such data collected over 220~PRs, we produce a SiV$^-$ number distribution, which closely follows a Poisson distribution with a mean $\bar{n}=0.53$ emitters per pillar [Fig.\,\ref{Fig:2Fab}(d)].
A certain discrepancy between the data and the Poissonian fit can be assigned to uncertainties in the experimental determination of the background signal.\\ 

In the following, we present a detailed characterization of the optical properties of individual SiV$^-$ at cryogenic conditions.
We employ a closed-cycle cold-finger cryostat to cool the diamond sample to $\sim7$\,K, where we conduct photoluminescence excitation (PLE) experiments.
For this, we tune a narrow-linewidth diode laser near resonance with the C transition of the SiV$^-$ [Fig.\,\ref{Fig:1Overview}(d)] and collect phonon sideband (PSB) emission as a function of excitation laser frequency. 

In Fig.\,\ref{Fig:1Overview}(e), we present the PLE spectrum of the narrowest linewidth that we observed and that exhibits a Lorentzian FWHM of $\Delta\nu=100.4\pm6.9$\,MHz. 
These data were obtained by averaging PLE spectra of fourteen successive laser sweeps across the C transition, followed by a Lorentzian fit (for details, see next section).
To benchmark the linewidth, we measure the optical lifetime of this SiV$^-$ by pulsed laser excitation at a wavelength of 515\,nm, followed by time-tagged ZPL photon collection [Fig.\,\ref{Fig:2Fab}(e), right panel]. 
An exponential fit to the photon decay trace yields a radiative lifetime $\tau_r=1.69\pm0.04$\,ns that corresponds to a lifetime-limited optical linewidth of $\Delta\nu=(2\pi\tau_r)^{-1}=93.9\pm2.2$\,MHz. 
 
To our knowledge, this is the first record of a lifetime-limited linewidth reported for $\lesssim50$\,nm shallow SiV$^-$ 
embedded in a diamond nanostructure, as required for nanoscale quantum sensing.
In addition, we find resonant saturation powers of this SiV$^-$ of $P_{\rm sat}=23.0\pm3.1$\,nW and a saturation count rate of $9.7\pm0.7$\,kcounts/s [Fig.\,\ref{Fig:2Fab}(e), left panel], typical for our devices.\\

\textbf{Charge repump-free photoluminescence excitation of SiV$^-$ centers in diamond}\\
PLE experiments with solid state emitters often require regular application of optical charge-resetting pulses using off-resonant laser light\,\cite{evansNarrowLinewidthHomogeneousOptical2016a,nicolasSubGHzLinewidthEnsembles2019,langLongOpticalCoherence2020,yurgensSpectrallyStableNitrogenvacancy2022}. Such charge resetting pulses are usually applied at wavelengths between 510 to 532 nm, to undo de-ionization events the emitter can undergo under resonant excitation. 
Importantly, such repumping perturbs the charge environment of the emitter and thus induces inhomogeneous broadening\,\cite{orphal-kobinOpticallyCoherentNitrogenVacancy2023,evansNarrowLinewidthHomogeneousOptical2016a,arjonamartinezPhotonicIndistinguishabilityTinVacancy2022}, precluding the observation of lifetime-limited optical linewidths. 
We refer to this measurement scheme as charge-repumped PLE (cr-PLE). 
Recently, Görlitz et al.\ \cite{gorlitzCoherenceChargeStabilised2022} have shown that exposing SnV$^-$ centers in diamond to 445\,nm laser light enables crf-PLE, reduces spectral diffusion and increases the brightness of SnV$^-$. Their charge-state lifetime (i.e. effective measurement time in crf-PLE) is, however, limited to about an hour under resonant excitation. They suggest that the same approach could also be beneficial to other group-IV vacancies, such as SiV$^-$.\\

 \begin{figure*}[ht]
    \centering
    \includegraphics[width=182mm]{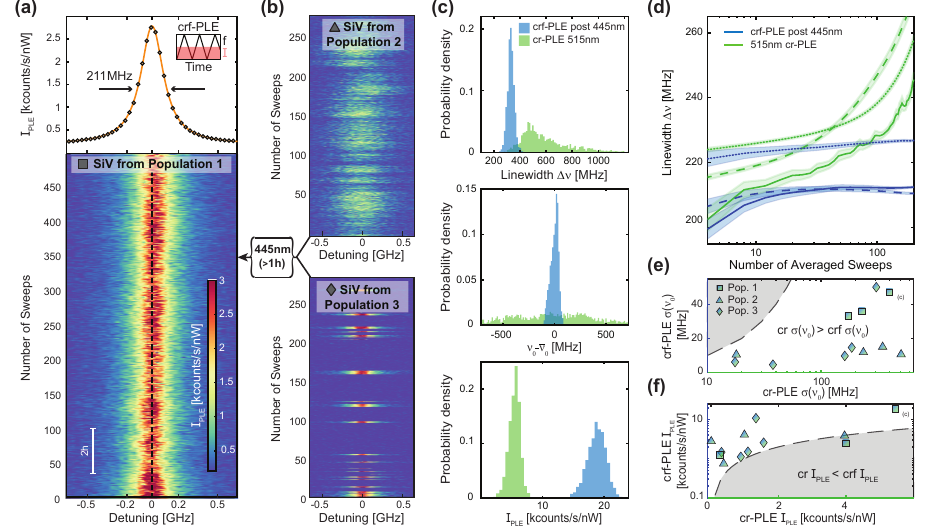}
    \caption{Repump-free photoluminescence excitation (crf-PLE) of SiV$^-$ in diamond parabolic reflectors and charge-state stabilization with 445\,nm laser light. 
(a) Top: crf-PLE measurement averaged over 500 single sweeps across the resonance while collecting (PSB) counts. 
A Lorentzian fit (yellow) to the data reports an inhomogenously broadened linewidth of $211.5\pm0.5$\,MHz, within a factor of 2.15 of the lifetime limit ($98.9\pm0.9$\,MHz) for this particular emitter. The inset illustrates how we conduct a crf-PLE experiment with constant laser intensity while modulating the laser frequency. 
Bottom: Trace of the PLE measurement, showing the 500 single sweeps. 
A dashed black line acts as a guide to the eye for zero detuning. 
(b) Diagram illustrating the the additional populations of SiV$^-$, classified by their behavior in crf-PLE (see main text). 
Populations 2 and 3 can be stabilized by applying high-powered ($>5$\,mW) 445\,nm laser light for extended durations ($>1$\,h). 
(c) Top panel: Histogram of single sweep Lorentzian linewidths $\Delta\nu$ of an SiV$^-$ initially in population 3, which was mapped to population 1. 
Green bars are results from 515\,nm charge-repump PLE (cr-PLE) and blue from crf-PLE after exposing the SiV$^-$ to the 445\,nm laser. 
This exemplary data set shows that 445\,nm exposure decreases single sweep linewidths compared to 515\,nm cr-PLE. 
Middle panel: Center frequency spread $\nu_0-\bar{\nu}_0$ extracted from the same PLE measurements, revealing that line stability is dramatically improved in crf-PLE. 
Bottom panel: Excitation power normalized peak intensity $\text{I}_{\text{PLE}}$ comparison for 515\,nm cr-PLE and crf-PLE. 
(d) Lorentzian fitted PLE linewidths $\Delta\nu$ as a function of the number of sweeps over which the PLE spectrum is averaged for three representative SiV$^-$, comparing standard 515\,nm cr-PLE (in green) and crf-PLE after 445\,nm (in blue) illumination, exemplifying the distinct behaviour between the measurement schemes. 
While crf-PLE converges to an average linewidth, linewidths measured during cr-PLE diverge [SI Fig.\,S8 for more data sets]. 
Data sets corresponding to the same SiV$^-$ are drawn with the same linestyle. The shaded area signifies the standard error. 
(e) Center frequency deviation $\sigma(\nu_0)$ comparison between crf-PLE on the y-axis and cr-PLE on the x-axis. The data point corresponding to the histograms in (c) is marked. 
A dashed black line denotes y = x. 
Markers denote the three populations introduced in (a) and (b). 
(f) Comparison between 515\,nm cr- and crf-PLE excitation power normalized peak intensity for the three populations as in (e), the dashed black line denoting y = x.}
 \label{fig:3PLE}
\end{figure*}
 
Our experiments on charge stabilization of SiV$^-$ with blue illumination revealed a similar, long-lasting effect. Prolonged (>$1$\,h) and high-intensity (>$5$\,mW) illumination of a PR with a 445\,nm laser led to persistently bright and stable PLE emission with narrow linewidths in crf-PLE, completely removing the need of charge resetting laser pulses, which is usually necessary for our samples.
An exemplary crf-PLE data set is depicted in Fig.\,\ref{fig:3PLE}(a), where the bottom (top) panel shows a sequence of 500 single crf-PLE sweeps (c.f.~SI Sec.~VI) and the corresponding average, respectively.
The data were continuously recorded over 14\,h, using exclusively near-resonant laser excitation.
The PLE resonance line retains its brightness and stability over the whole measurement duration and yields an averaged, inhomogeneously broadened linewidth of $\Delta\nu=211.5\pm0.5$\,MHz, within a factor of 2.15 of its lifetime limit of $98.9\pm0.9$\,MHz, which we evaluated by an independent excited state lifetime measurement at $7$\,K.\\

Having observed the positive impact of 445\,nm illumination on the optical properties of SiV$^-$ in resonant excitation, resulting in crf-PLE with improved linewidths and stability, we further address the reproducibility and effectiveness of this phenomenon. To do so, we investigated twelve PRs, nine of which contain single SiV$^-$, with the following measurement sequence: 
For each SiV$^-$, and before exposing them to anything other than 515\,nm laser light, we start by performing crf-PLE to assess the initial charge stability and linewidth.
During these measurements, we have observed three clearly distinct, roughly equally distributed SiV$^-$ populations, classified by their behaviour in crf-PLE: SiV$^-$ in population 1 exhibit continuous, stable and bright emission with narrow linewidths [Fig.\,\ref{fig:3PLE}(a)]; SiV$^-$ in population 2 present dimmer emission with large spectral diffusion and broader single sweep linewidths under continuous resonant excitation [Fig.\,\ref{fig:3PLE}(b) - top]; SiV$^-$ in population 3 show charge state instabilities (blinking), where it is not possible to perform continuous crf-PLE [Fig.\,\ref{fig:3PLE}(b) - bottom].
Secondly, we perform `traditional' cr-PLE using a 515\,nm charge repump laser to benchmark the PLE linewidth under this measurement scheme. 
Lastly, we continuously expose the SiV to 445\,nm excitation at $>5$\,mW for prolonged periods of time ($>1$\,h), and repeat the crf-PLE linewidth measurement.
From this measurement series, we found above all that SiV$^-$ initially in population 2 and 3 can be mapped to population 1 after prolonged $445$\,nm laser exposure, leading to drastically improved properties.

An exemplary result of the outcome of the above-mentioned measurement series performed on a single SiV$^-$ initially from population 3 is presented in Fig.\,\ref{fig:3PLE}(c).
The histograms show the probability of single sweep Lorentzian linewidths $\Delta\nu$ (top panel), center frequency spread $\nu_0-\bar{\nu}_0$ (middle panel) and excitation power normalized peak intensity $\text{I}_{\text{PLE}}$ (bottom panel) for cr-PLE and crf-PLE. The data clearly show how crf-PLE after 445\,nm laser exposure yields both a strongly reduced single sweep linewidth and center frequency spread, and a highly increased peak intensity compared to 515\,nm cr-PLE. The results we obtained in this way for the twelve investigated SiV$^-$ are summarized in Fig.\,\ref{fig:3PLE}(d)-(f). 
In 3(d), we show three representative data sets of PLE linewidths measured on single SiV$^-$ as a function of the number of single sweeps over which the data were averaged. We compare the result for 515 nm cr-PLE (green) and crf-PLE after illuminating their respective PRs with the 445 laser (blue) [SI Sec. VI for more data sets].
While after a few tens of sweeps, crf-PLE converges to a linewidth in the range of $\sim200$\,MHz, the averaged cr-PLE linewidths diverge as a function of the number of averaged sweeps. 
These data are testament to the absence of excess spectral diffusion, i.e.\ spectral wandering\,\cite{wolfowiczQuantumGuidelinesSolidstate2021} is completely eliminated when performing PLE without a charge repump laser, which enables long-time measurements without loss of optical coherence. 
Lastly, in order to further compare the two protocols, for each SiV$^-$ we extracted the center frequency standard deviation $\sigma(\nu_0)$ and the average power normalized peak intensity I$_{\text{PLE}}$ from all single sweep crf-PLE spectra, and plot them against the corresponding values for the same SiV$^-$ under 515\,nm cr-PLE (Fig.\,\ref{fig:3PLE}(e) and (f)).

We consistently find that in crf-PLE, SiV$^-$ exhibit higher peak PLE intensities, less background fluorescence, and display strikingly reduced linewidths and spectral diffusion.
However, it is noteworthy that for some SiV$^-$ initially in population 1, illumination with 445\,nm light slightly decreases their brightness and stability [SI Sec.~VI].
We confirmed that this beneficial effect of prolonged 445\,nm laser illumination is persistent and neither specific to a diamond sample nor the Si implantation dose.
Specifically, we confirmed the same behaviour as presented in Fig.\,\ref{fig:3PLE} for SiV$^-$ in a second sample B, which was prepared in the same way as the first sample, but with a $^{28}$Si implantation dose five times higher
(as a sole, slight difference between the two, we found that in the second sample, Population 3 made up for a smaller percentage than in the low-dose sample A).
For both samples, the beneficial effect of 445\,nm laser exposure persisted throughout the timescale of this study (several months) and was neither affected by continuous, high-power laser illumination (be it resonant or off-resonant), by long idle times in the dark, nor by thermal cycling of the samples.
These combined findings suggest that illumination of 445\,nm laser light is a generally applicable means for permanently stabilizing the charge-environment of shallow SiV$^-$ centers, such that coherent optical addressing can be performed without any further charge repumping.

Only in the case where such continuous crf-PLE measurements can be conducted on a given SiV$^-$ from the beginning (population 1), 445\,nm laser illumination might slightly deteriorate the SiV$^-$ optical properties and calls for cautious use of blue laser illumination.\\


\begin{figure}[ht]
    \centering
    \includegraphics[width=82.55mm]{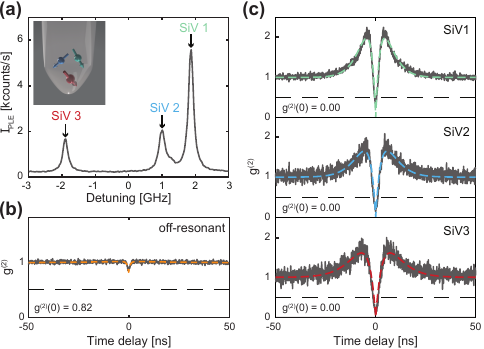}
    \caption{Addressing spectrally distinct individual SiV$^-$ hosted in the same nanostructure. 
(a) crf-PLE measurement revealing three resonances, which we attribute to C transitions of three distinct SiV$^-$ within the same parabolic reflector (PR). 
The inset illustrates such a situation. 
(b) Off-resonant $g^{(2)}(\tau)$ on the PR in question, showing that indeed, more than one emitter is being addressed. 
(c) Resonant $g^{(2)}(\tau)$ of each individual resonance shown in (a) obtained by subsequently tuning the 737\,nm laser into each resonance and collecting (PSB) photons. 
These data show the ability to address one individual emitter in a multi-emitter nanostructure, increasing the yield of scanning probes in a sample.}
 \label{Fig:4MultipleSiV$^-$s}
\end{figure}

\textbf{Addressing spectrally distinct individual SiV$^-$ in multi-SiV nanostructures}\\
The narrow and stable SiV$^-$ linewidths we demonstrated enable the addressing of individual emitters in nanostructures that contain multiple, spectrally distinct SiV$^-$.
Such a situation is illustrated in Fig.\,\ref{Fig:4MultipleSiV$^-$s}(a), where a crf-PLE measurement after prolonged 445\,nm laser illumination shows three PLE resonances, which we attribute to the C transitions of three separate SiV$^-$ hosted in a single PR on the high-density diamond sample.
The shift in their transition frequency likely results from local variations in strain or electric field in the surroundings of each SiV$^-$. A $g^{(2)}(\tau)$ measurement conducted under off-resonant optical excitation at $515$\,nm [Fig.\,\ref{Fig:4MultipleSiV$^-$s}(b)] reveals a value $g^{(2)}(0)=0.82\pm0.01$, which indicates that indeed more than one emitter is present in this particular PR. Since however, owing to our charge stabilization protocol, the resonances of the three SiV$^-$ remain spectrally distinct, one can individually address each SiV$^-$ despite their localization in a nanoscale volume.
We demonstrate this by resonantly driving each of the three SiV$^-$ and recording a corresponding $g^{(2)}(\tau)$ trace. Indeed, the three photon autocorrelation traces [Fig.\,\ref{Fig:4MultipleSiV$^-$s}(c)] all show values of $g^{(2)}(0)$ close to zero ($g^{(2)}(0)=0.00\pm0.03$, $0.00\pm0.03$, $0.00\pm0.03$, respectively), indicating that only one SiV$^-$ at a time is being optically excited in this case. 
For nanoscale quantum sensing with SiV$^-$, this result brings the interesting perspective of performing single-spin sensing in nanostructures containing small ensembles of spins, which would find immediate applications, for example, in covariance magnetometry\,\cite{rovnyNanoscaleCovarianceMagnetometry2022}.

\section*{\label{sec:Discussion}Discussion}
In this work, we demonstrated the robust and reproducible creation of single narrow-linewidth SiV$^-$ color centers located within a few tens of nanometers from the end facets of individual diamond nanopillars. 

Nearly all SiV$^-$ investigated here display inhomogenously broadened linewidths within a factor of two from the lifetime limit in crf-PLE over long timescales, and single sweep linewidths that, at times, approach their respective lifetime limit.
These results are enabled by a combination of a high temperature vacuum annealing step introduced after nanopillar fabrication, 
and the application of a novel, optical charge stabilization protocol based on extended, single-time exposure of SiV$^-$ to continuous-wave 445\,nm laser light.
The latter permanently and entirely removes the need for charge repumping in resonant excitation experiments and yields improvement in several key figures of merit of resonant optical excitation of SiV$^-$ centers. 
Specifically, the optical charge stabilization leads to enhanced brightness, reduced spectral diffusion, and charge state preservation for those SiV$^-$ which suffered from de-ionization under resonant excitation.
While the microscopic origins underlying the demonstrated optical charge stabilization scheme remain unexplained and depletion of the charge environment may play a role\,\cite{andersonElectricalOpticalControl2019}, we anticipate that our results will trigger significant further research in theory and experiment.

Our results constitute a major step towards the use of SiV$^-$ as nanoscale quantum sensors for applications under extreme conditions, such as single spin scanning magnetometry at sub-kelvin temperatures and tesla-range magnetic fields\,\cite{fuHighSensitivityMomentMagnetometry2020}. 
Furthermore, our easy-to-implement charge-stabilization scheme will find immediate applications in other quantum technology applications of SiV$^-$, including the development of quantum repeaters\,\cite{bayerQuantumRepeaterPlatform2022}, quantum networks\,\cite{sipahigilIntegratedDiamondNanophotonics2016b,nguyenQuantumNetworkNodes2019} or indistinguishable single photon sources\,\cite{sipahigilIndistinguishablePhotonsSeparated2014}. 
Lastly, it is conceivable that our approach for generating and stabilizing near-surface color centers with high optical coherence extends to other color centers in diamond or in other wide-bandgap hosts such as hBN\,\cite{grollControllingPhotoluminescenceSpectra2021} or SiC\,\cite{nagyHighfidelitySpinOptical2019}.\\

\section*{Methods}

\small\textbf{Optical Setups}

To characterize our diamond samples at room temperature (RT), we employ a home-built confocal microscopy setup. This setup is equipped with a continuous wave (cw) 515\,nm diode laser (Cobolt 06-MLD) for off-resonant excitation. At low temperature ($\sim7$\,K), the sample is housed in a variable-temperature closed-cycle cryostat (attocube attoDRY800) and we perform optical spectroscopy again with a home-built confocal microscopy setup. For this setup, we put to use cw 445\,nm and cw 515\,nm diode lasers (Cobolt 06-MLD) for off-resonant excitation as well as a narrow-linewidth (200\,kHz) tunable (720-739\,nm) diode laser (Sacher Lasertechnik LION) for resonant excitation. Frequency modulation and control of the LION is achieved through a PID loop of a HighFinesse WS-U wavelength meter. Additionally, we stabilize the intensity of the resonant laser through the PID loop of a home-built optical intensity stabilization (OIS) module (Physics Basel SP 999). In case of the resonant laser and if applicable to the experiment in question, we carve pulses with an acousto-optical modulator (AOM, G\&H R15210), whereas the off-resonant lasers can be electronically modulated out of the box. For both RT and LT, we focus laser light onto the sample with a 0.8 NA objective (Olympus LMPLFLN 100X) thermalized at RT, through which we also collect the signal. Photons are detected by silicon-based single-photon counting modules (SPCM, Excelitas AQRH-33-FC). We perform photon-correlation measurements using a time-correlated single-photon counting (TCSPC) system (PicoQuant PicoHarp 300) connected to the SPCMs. For optical lifetime measurements we use a pulsed laser (NKT Photonics SuperK Extreme EXW-12) tuned to 515\,nm and said TCSPC system. Optical spectra are recorded on the camera of a spectrograph (Princeton Instruments SP-2500).\\

\textbf{Data availability}\\
The data that support the findings of this study are available from the corresponding author upon reasonable request.

\section*{Acknowledgements}

We gratefully acknowledge Christoph Becher and Dennis Herrmann for fruitful discussions as well as Silvia Ruffieux for helping to fabricate PRs on sample A. 
We further acknowledge financial support through QuantERA project ``sensExtreme'' (Grant No.~$205573$), from the Swiss Nanoscience Institute, and through the Swiss NSF Project Grant No.~$188521$.

\section*{Author information}
These authors contributed equally: Josh A. Zuber, Minghao Li.\\

\textbf{Authors and Affiliations}\\
\textbf{Department of Physics, University of Basel, Klingelbergstrasse 82, Basel, CH-4056, Switzerland}\\
J. A. Zuber, M. Li, M. Grimau Puigibert, J. Happacher, P. Reiser, B. J. Shields \& P. Maletinsky\\

\textbf{Swiss Nanoscience Institute, Klingelbergstrasse 82, Basel, CH-4056, Switzerland}\\
J. A. Zuber \& P. Maletinsky\\

\textbf{Contributions}\\
All experiments, data analysis and sample preparation shown here were conducted by J.A.Z. and M.L. under the supervision of P.M.. M.G.P. started the project under P.M.'s supervision, built the first setup and fabricated preliminary samples. M.G.P., together with J.H., conducted preliminary studies on and fabricated Sample B. P.R. fabricated PRs on sample A with the help of J.A.Z. and M.L.. B.J.S. fabricated PRs on preliminary samples. J.A.Z., M.L. and P.M. wrote the manuscript. All authors discussed the data and commented on the manuscript.\\

\textbf{Competing interests}\\
The Authors declare no competing interests.\\

\textbf{Correspondence} and requests for materials should be addressed to P. Maletinsky

\newpage

\bibliographystyle{naturemag}
\bibliography{Main_shallow_narrow_SiVs_in_pillars.bib}

\end{document}



\title{Supplementary Information for:\\Shallow Silicon Vacancy Centers with lifetime-limited optical linewidths in Diamond Nanostructures}
\author{Josh A. Zuber}
\thanks{These authors contributed equally.} 
\affiliation{Department of Physics, University of Basel, CH-4056 Basel, Switzerland}
\affiliation{Swiss Nanoscience Institute, University of Basel, CH-4056 Basel, Switzerland}
\author{Minghao Li}
\thanks{These authors contributed equally.} 
\affiliation{Department of Physics, University of Basel, CH-4056 Basel, Switzerland}
\author{Marcel.li Grimau Puigibert}
\affiliation{Department of Physics, University of Basel, CH-4056 Basel, Switzerland}
\author{Jodok Happacher}
\affiliation{Department of Physics, University of Basel, CH-4056 Basel, Switzerland}
\author{Patrick Reiser}
\affiliation{Department of Physics, University of Basel, CH-4056 Basel, Switzerland}
\author{Brendan J. Shields}
\affiliation{Department of Physics, University of Basel, CH-4056 Basel, Switzerland}
\author{Patrick Maletinsky}
\email{patrick.maletinsky@unibas.ch}
\affiliation{Department of Physics, University of Basel, CH-4056 Basel, Switzerland}
\affiliation{Swiss Nanoscience Institute, University of Basel, CH-4056 Basel, Switzerland}

\date{July 24, 2023}

\maketitle


\section{\label{SISec:Yield}SiV$^-$ Yield Estimation}

We predicted our target implantation density of \qty{6e9} ions/cm$^2$ by using yield estimates from previous samples, where we observed a Si $\rightarrow$ SiV$^-$ conversion efficiency of $\sim0.03$ after implantation with $^{28}$Si ions at a fluence of \qty{1.5e11} ions/cm$^2$ and 80\,\unit{\kilo\electronvolt}, the first annealing procedure and parabolic reflector fabrication and of $\sim0.18$ after the second anneal. 

After the first annealing and after parabolic reflector (PR) fabrication, we could still find single SiV$^-$ centers in our PRs even at this high implantation fluence, which was not the case after the second anneal. Based on simple back-of-the envelope calculations, we estimated that a fluence of \qty{1e10} ions/cm$^2$ would yield single SiV$^-$ centers in PRs with diameters around 300\,\unit{\nano\meter} in combination with two annealing procedures. We then opted for the slightly lower fluence to facilitate subsequent characterization. The reason for the switch from $^{28}$Si to $^{29}$Si
was based on technicalities with the implantation provider. If a single stage implanter is used, it is possible that $^{28}$N$_2$ is implanted alongside $^{28}$Si, leading to unwanted contamination of the sample and an increased yield of NV centers. We do not expect an impact on optical coherence based on this change.

\section{\label{sec:SRIM}Stopping Range of Ions in Matter (SRIM) Simulation}

We estimate the depth of our emitters based on the Stopping Range of Ions in Matter (SRIM) Monte-Carlo simulations. The parameters for the simulation are as follows. For the ions, we select Si and choose a mass of 28.977\,amu, an energy of 80\,\unit{\kilo\electronvolt} and an angle of incidence of 7\,\unit{\degree}. For the diamond target, we select C with the density of diamond of 3.53\,\unit{\gram\per\cm\cubed}. The stopping range of the $^{29}$Si is estimated to be $54.3\pm13.9$\,\unit{\nano\meter}. We also note that due to our two high temperature annealing steps, we expect some graphitization of the surface to occur; this graphitization layer is removed by tri-acid cleaning, bringing the emitters in our samples even closer to the surface.
\begin{figure}[h]
    \centering
    \includegraphics[width=86mm]{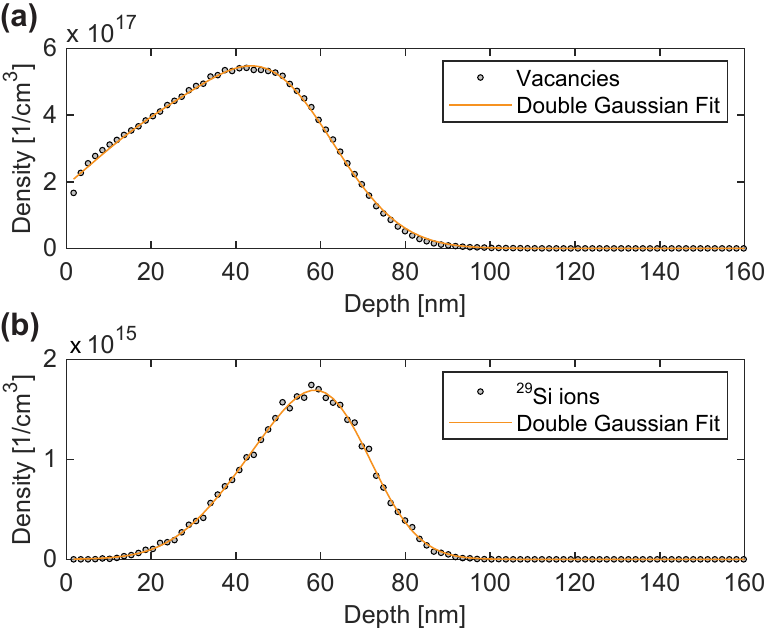}
    \caption{SRIM results for $^{29}$Si implantation into diamond at an angle of 7\,\unit{\degree} and 80\,\unit{\kilo\electronvolt}. (a) The resulting density distribution for the vacancies as a function of depth in the diamond. Fitted with double Gaussians (yellow line) in order to calculate the most likely depth of the vacancies of 44.2\,\unit{\nano\meter}. (b) Resulting density distribution for the $^{29}$Si ions as a function, again fitted with a double gaussian, confirming the SRIM prediction of an average depth of $54.3\pm13.9$\,\unit{\nano\meter} for the $^{29}$Si ions.}
 \label{SI_Fig:SRIM}
\end{figure}

\section{\label{Sec:RTCharac}Room Temperature PR field characterization}

\begin{figure}[h]
    \centering
    \includegraphics[width=0.8\textwidth]{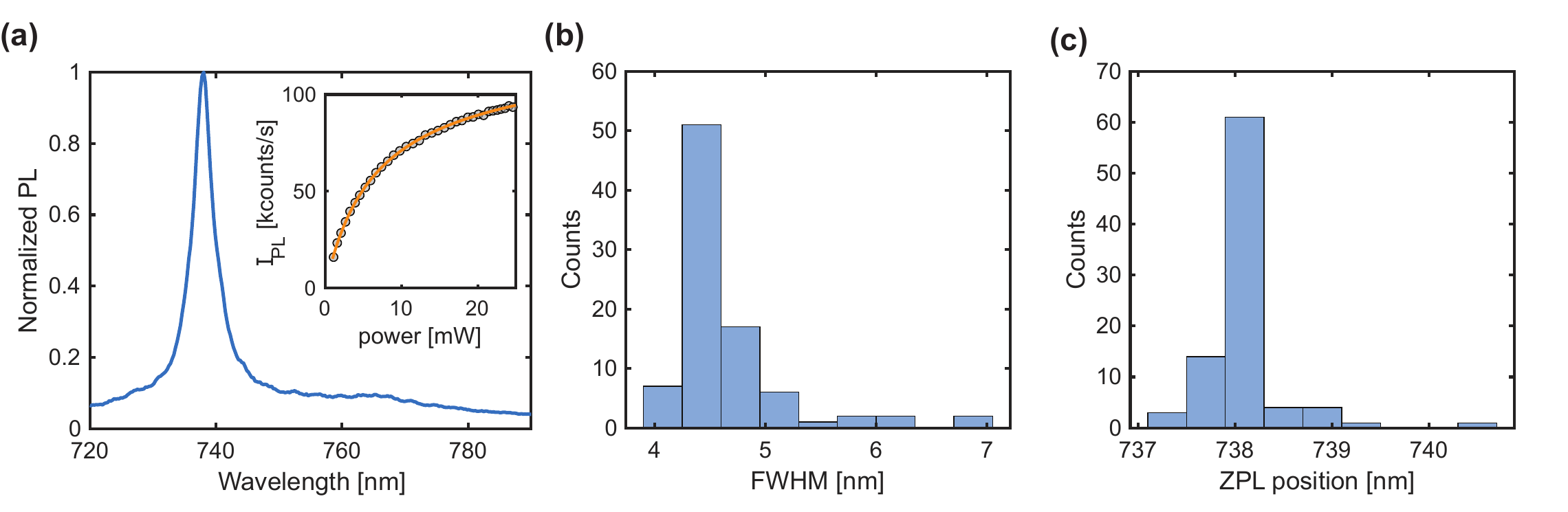}
    \caption{(a) Example of a typical room temperature photoluminescence (PL) spectrum of the SiV$^-$ center, the zero-phonon line (ZPL) of SiV$^-$ centered at around 738\,\unit{\nano\meter}. The inset shows the room temperature saturation curve when collecting ZPL photons upon off-resonant excitation at 515\,\unit{\nano\meter} (gray scatters) with the fitting (yellow line). (b) and (c) Histograms of the linewidth (FWHM) and the ZPL position, obtained by fitting the room temperature spectra with a Lorentzian, of in total 88 pillars that contain detectable SiV$^-$ centers from sample A.}
 \label{SI_Fig:RT_spect}
\end{figure}

\begin{figure}[h]
    \centering
    \includegraphics[width=0.57\textwidth]{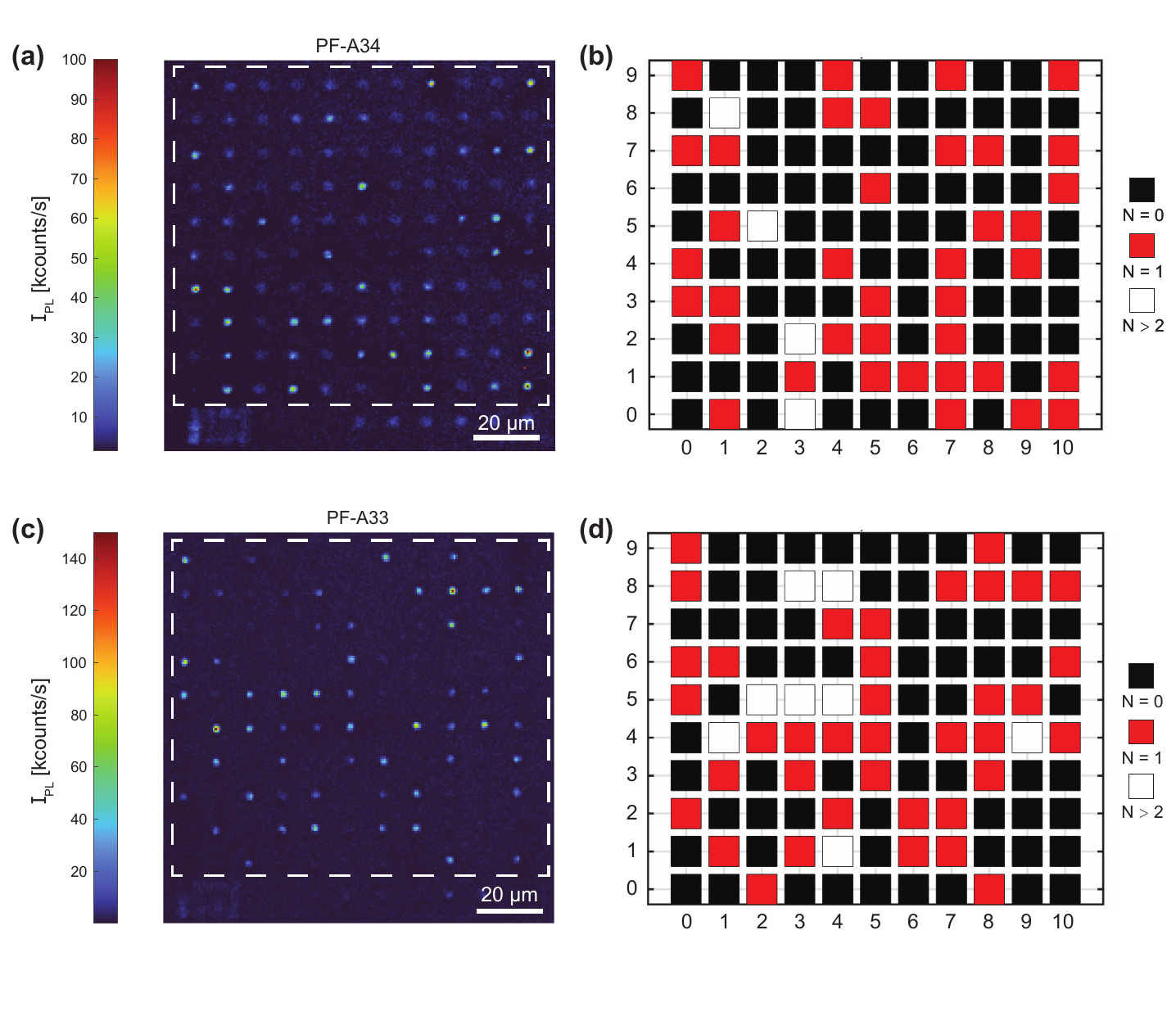}
    \caption{(a) and (c) Confocal scans of the two PR arrays where we perform the systematic characterization (optical spectra, saturation curve, lifetime and $g^{(2)}(\tau)$ measurements as mentioned in the main text) at room temperature (PRs inside the white dashed box). (b) and (d) respectively present the result of the number of emitters embedded in the PRs at the corresponding coordinate, facilitating recognition of individual PRs when the sample is transferred to the cryostat for low temperature characterizations. The histogram presented in Fig. 2(d) in the main text is based on these results.}
 \label{SI_Fig:RT_scans}
\end{figure}

The room temperature (RT) PR characterizations shown in the main text are performed on 220 PRs investigated in two array fields (confocal scans shown in Fig. \ref{SI_Fig:RT_scans}(a) and (c)). As the arrays are regularly fabricated, a series of systematic measurements can be easily performed on each PR. For every PR in the selected array, a ZPL spectrum measurement is conducted to filter out the empty PR that contain no detectable SiV$^-$. Then, for every non-empty PR, a saturation curve, $g^{(2)}(\tau)$ and RT excited state lifetime measurement are carried out successively. An exemplary RT ZPL spectrum under off-resonant excitation with a $515~$nm laser is shown in Fig. \ref{SI_Fig:RT_spect} (a) with the associated saturation curve shown in its inset. Fig. \ref{SI_Fig:RT_spect} (b) and (c) present the histograms of the ZPL linewidths and position for non-empty PRs. These results clearly show that the SiV$^-$ centers we created in PRs are highly homogeneous and mostly unstrained.
Fig. \ref{SI_Fig:RT_scans} (b) and (d) shows the number of SiV in the corresponding PR in the confocal scan in Fig. \ref{SI_Fig:RT_scans} (a) and (c) obtained by the systematic $g^{(2)}(\tau)$ measurements.

\section{\label{SiSec:g2corr}$g^{(2)}(\tau)$ Background Correction}

For the background correction of second order correlation function $g^{2}(\tau)$ data, we adopt the method introduced by Brouri et al. \cite{brouriPhotonAntibunchingFluorescence2000}. 
The raw data $W(\tau)$ is obtained in the histogramming mode of the PicoHarp 300 by measuring the waiting time between two photon detectors in the Hanbury Brown and Twiss (HBT) setup. Since we operate in the low count rate regime, the histogram serves as a reliable approximation of the exact photon correlation obtained through full counting statistics in time tagger mode. Therefore, we normalize the raw data $W(\tau)$ by dividing it by its value at a large time delay ($\tau\gg\tau_c$) where $W(\tau)$ tends towards a constant value. 
\begin{equation}
    g^{(2)}_{norm}(\tau) = \frac{W(\tau)}{W(\tau\gg\tau_c)}
\end{equation}

This normalization method is validated by comparing the value of $W(\tau\gg\tau_c)$ with the coincidence number $C$ obtained from an ideal Poissonian photon source:

\begin{equation}\label{eq:g2norm}
   C = N_1N_2\Delta t T,
\end{equation}

where $N_1,N_2$ are the count rates on detectors 1 and 2 respectively, $\Delta t$ is the selected time bin width and $T$ is the total acquisition time. In our measurement, although we do not have access to exact $N_1,N_2$ during the acquisition, the current count rate on each detector is saved every five minutes. We found that in most of the cases, $C\sim W(\tau\gg\tau_c)$.

After normalizing the raw data, we further correct for the background by the signal-to-background ratio $\rho(P)=S(P)/(S(P)+B(P))$ for a given power $P$. This value is determined individually for each SiV$^-$ center by fitting the saturation curve and extracting the values of S(P) and B(P) from the fit at the laser intensity used to acquire the corresponding $W(\tau)$. The background corrected $g^{(2)}_{\text{corr}}(\tau)$ is then given by

\begin{equation}
   g^{(2)}_{\text{corr}}(\tau) = \frac{g^{(2)}_{\text{norm}}(\tau)-(1-\rho(P)^2)}{\rho(P)^2}.
\end{equation}



\section{cr-PLE with 515\,\unit{\nano\meter} and 445\,\unit{\nano\meter} repump pulses}

\begin{figure}[!h]
    \centering
    \includegraphics[width=180mm]{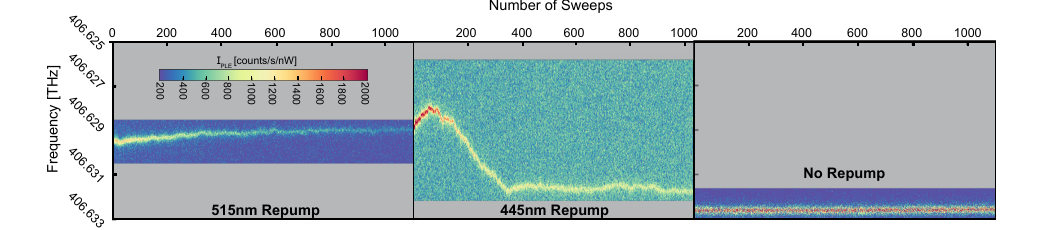}
    \caption{PLE traces for different repump schemes on the same PR. Left panel: For 515\,\unit{\nano\meter} cr-PLE, we observe spectral diffusion on the order of below 1\,\unit{\giga\hertz}, while for 445\,\unit{\nano\meter} cr-PLE (middle panel), spectral diffusion first increases to around 2\,\unit{\giga\hertz} before saturating. Finally, in a crf-PLE (right panel), spectral diffusion is minimized, albeit the resonance shifts slightly from the saturated frequency observed in the 445\,\unit{\nano\meter} cr-PLE. These experiments were performed one after the other with some time in the dark between each measurement. A sweep takes around 26\,\unit{\second} for the first and third measurement and around 64\,\unit{\second} for the second measurement, owing to its larger range.}
 \label{SI_Fig:PLETraces}
\end{figure}

\begin{figure}[!h]
    \centering
    \includegraphics[width=140mm]{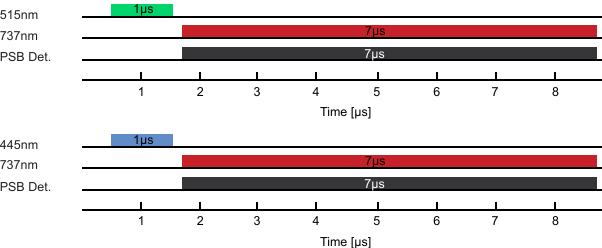}
    \caption{Exact pulse sequences for cr-PLE measurements with 515\,\unit{\nano\meter} and 445\,\unit{\nano\meter} respectively. Off-resonant repump pulses have a length of 1\,\unit{\micro\second} whereas the resonant laser pulse and the detection window have a length of 7\,\unit{\micro\second}} 
 \label{SI_Fig:PulsSeq}
\end{figure}

\newpage

\section{\label{SiSec:PLEStatAllData}PLE Statistics and Single Sweep discussion}

\begin{figure}[h]
    \centering
    \includegraphics[width=1\textwidth]{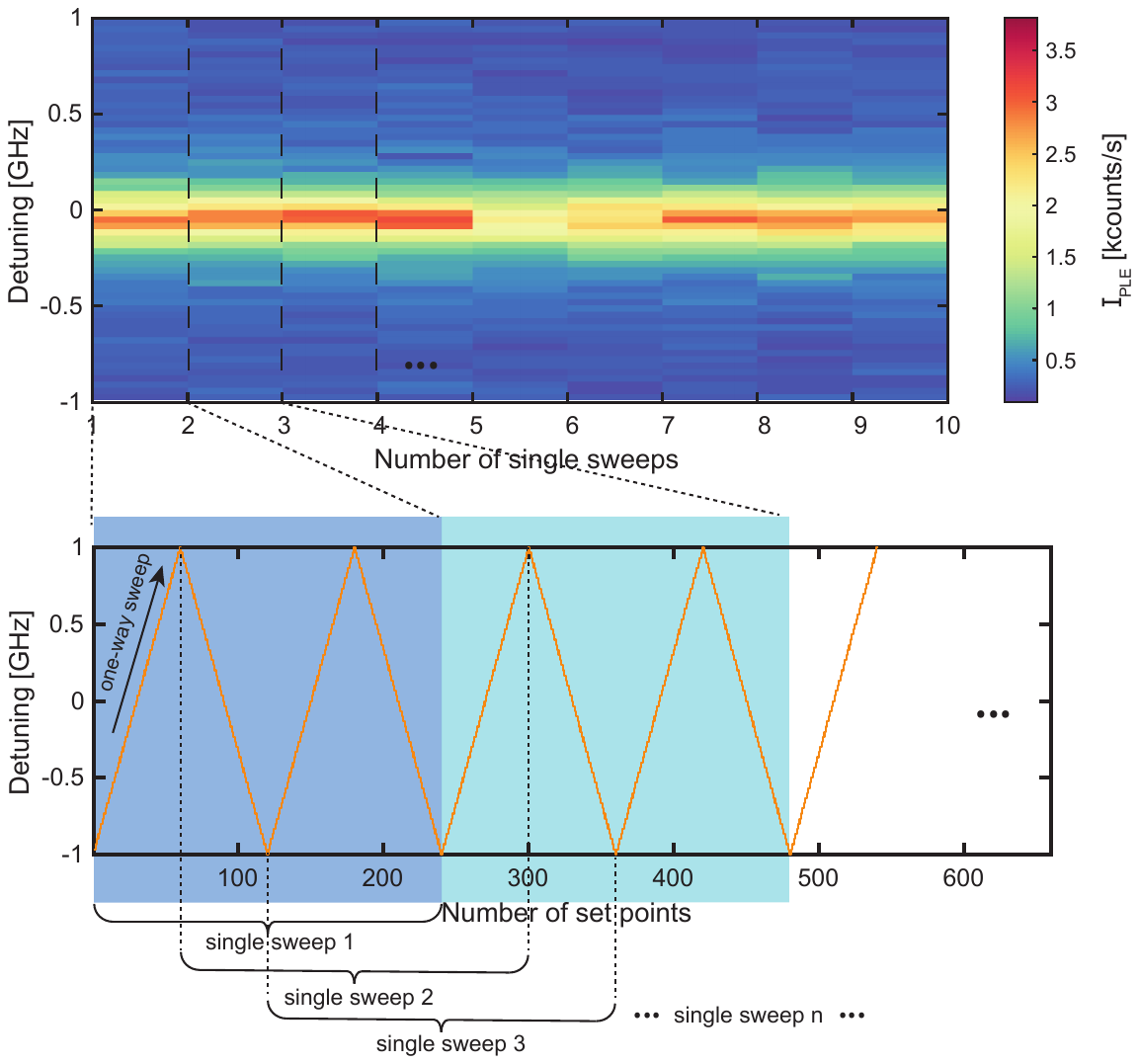}
    \caption{The upper panel shows an exemplary PLE trace with a total number of 9 single sweeps with a sweeping range from -1 GHz to 1 GHz. One single sweep PLE spectrum in this trace consists of two entire actual back-and-forth laser sweeps from -1 GHz to 1 GHz as sketched in the lower panel. As one single sweep data set in the trace is composed of four one-way laser sweeps across the transition, the single sweeps can be sampled in an overlapping scheme as shown in the lower panel.}
 \label{SI_Fig:sweep}
\end{figure}

The statistics of PLE spectral quantities (linewidth, peak position and peak intensity) presented in the main text are extracted from so-called single sweeps along a long PLE measurement. In this section, we provide a detailed description of the actual laser sweeps in our PLE measurements and how we define the term "single sweep". The actual laser (forward-) sweep is achieved by sequentially adjusting the set point from the initial to the final detuning frequency (or vice-versa, respectively, for the backwards sweep), with each set point regulated by a PID loop via a wavemeter (HighFinesse WS-U). The dwell time for the laser at each set point is set to be $\sim0.2~$s allowing the PID loop sufficient time to reach the set point, while PL intensity is measured by an APD with an integration time of $20~$ms. The resulting back-and-forth actual laser sweeps possess a saw-tooth shape as shown in the lower panel of Fig. \ref{SI_Fig:sweep}. A typical number of set points we choose is 61 points for a one-way laser sweep across a detuning range from $-1~$GHz to $1~$GHz. The total time for one back-and-forth laser sweep is $\sim 26~$s. In order to strike a balance between a satisfactory signal-to-noise ratio in single sweep PLE spectra and ensuring the visibility of spectral diffusion dynamics, we define here two complete back-and-forth actual laser sweeps (or four one-way laser sweeps) to be one "single sweep" as illustrated in Fig. \ref{SI_Fig:sweep}, i.e. we average the data over these four actual laser sweeps to produce what we call a "single sweep". 
When analyzing the statistics of PLE spectral quantities of a single sweep, we adopt the overlapping sampling scheme outlined in the lower panel of Fig. \ref{SI_Fig:sweep}, where for a total number of actual laser one-way sweeps $N$, the number of samples for a single sweep would be $N-3$. However, when displaying a PLE trace as shown in upper panel of Fig. \ref{SI_Fig:sweep}, the successive four one-way sweeps compose the single sweep in the trace.

\begin{figure}[h]
    \centering
    \includegraphics[width=130mm]{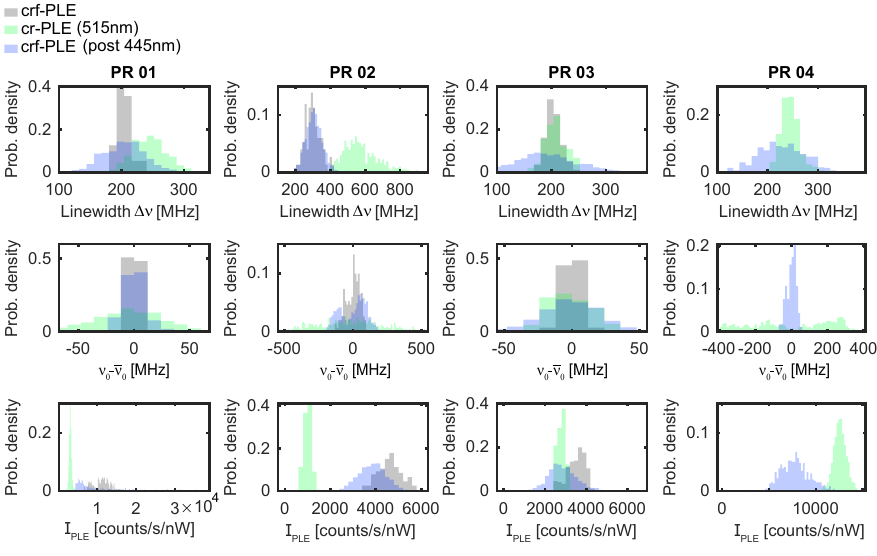}
 \label{SI_Fig:Histograms1}
\end{figure}
\vspace{-4cm}
\begin{figure}[h]
    \centering
    \includegraphics[width=130mm]{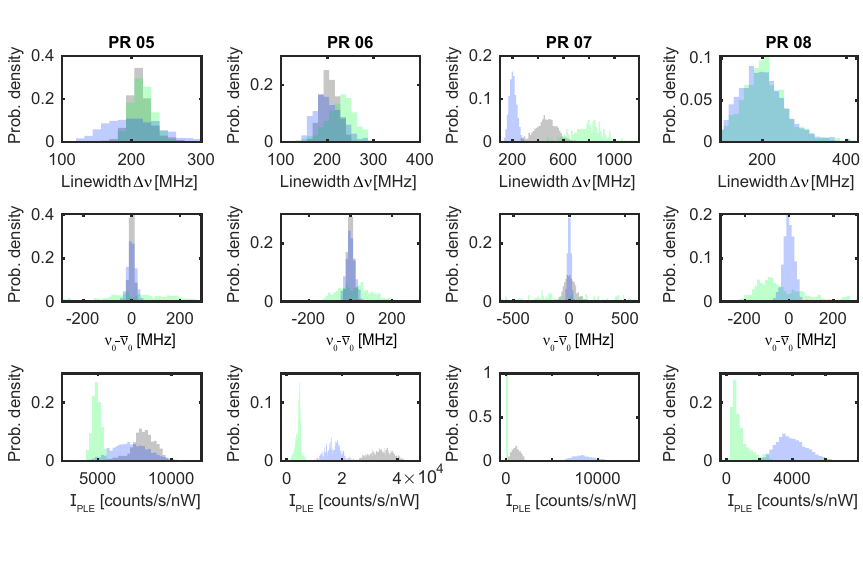}
 \label{SI_Fig:Histograms2}
\end{figure}
\vspace{-4cm}
\begin{figure}[h]
    \centering
    \includegraphics[width=130mm]{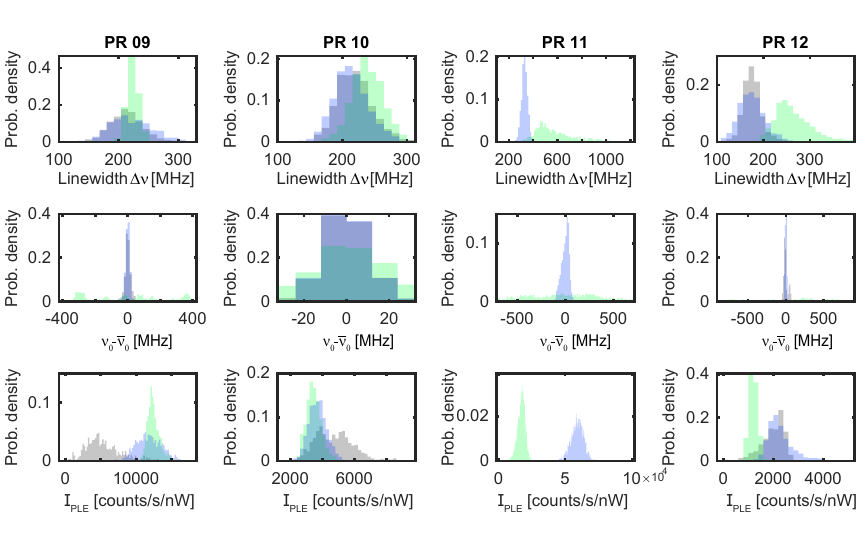}
    \caption{Single sweep linewidths $\Delta\nu$, center frequency spread $\nu_0-\bar{\nu}_0$ and peak intensity I$_{\text{PLE}}$ histograms for 12 PRs, comparing crf-PLE before 445\,\unit{\nano\meter} exposure (grey, if possible for the defect center in question), 515\,\unit{\nano\meter} cr-PLE (green) and crf-PLE after 445\,\unit{\nano\meter} exposure (blue).}
 \label{SI_Fig:Histograms3}
\end{figure}

\begin{figure}[h]
    \centering
    \includegraphics[width=120mm]{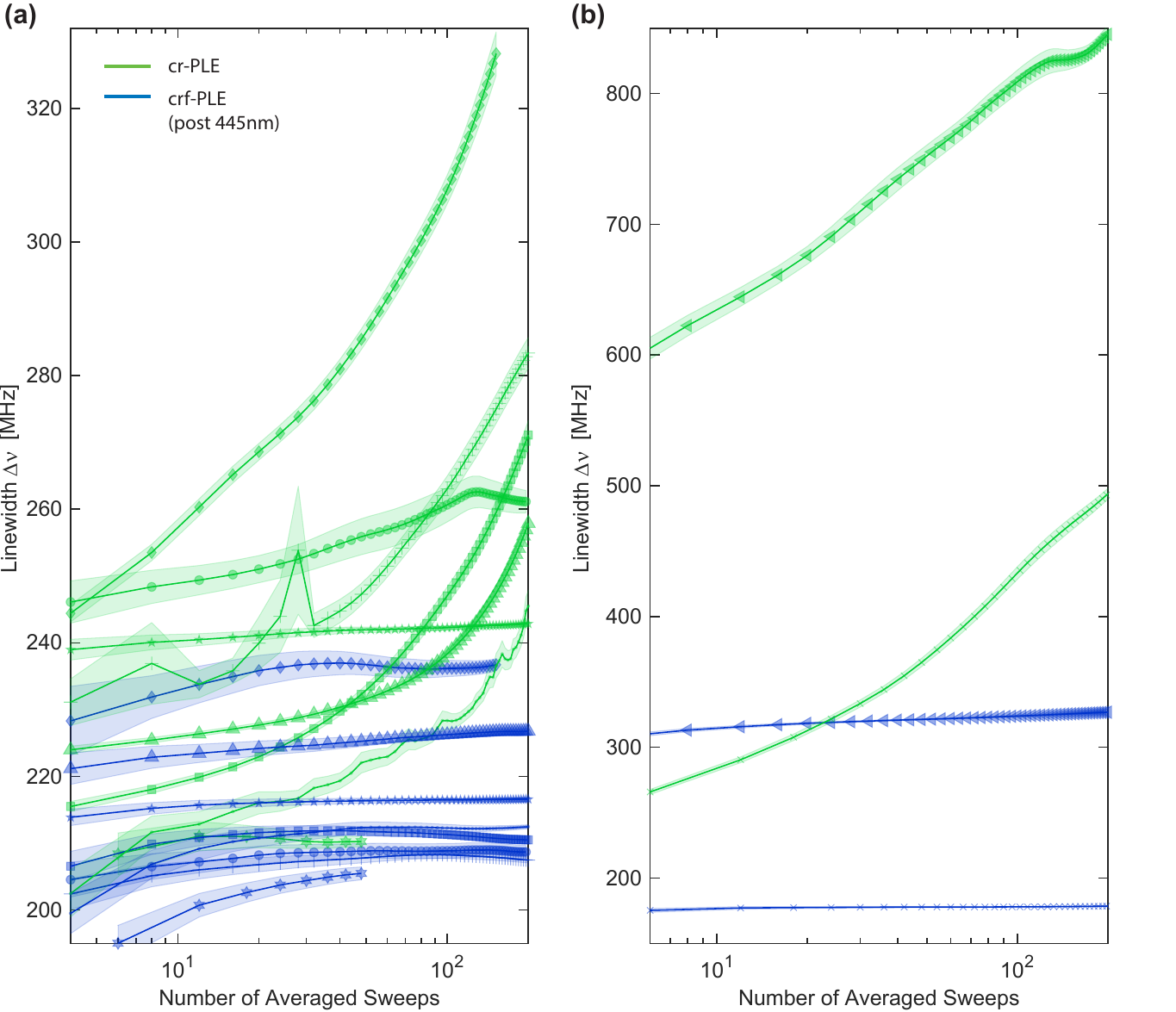}
    \caption{Linewidth $\Delta\nu$ evolution as a function of the number of averaged sweeps, additional data sets. (a) $\Delta\nu$ evolution under crf-PLE (blue) and cr-PLE with 515\,\unit{\nano\meter} repump pulses (green) for PR 01, PR 03-06, PR 08-10. Symbols used correspond to the same PR. The shaded area denotes the standard error. (b) $\Delta\nu$ evolution under crf-PLE (blue) and cr-PLE with 515\,\unit{\nano\meter} repump pulses (green) for PR 02, PR 07. Symbols used correspond to the same PR. The shaded area denotes the standard error.}
 \label{SI_Fig:LWEvolution}
\end{figure}

\section{\label{SiSec:PowerBroad}Power Broadening}

\begin{figure}[h]
    \centering
    \includegraphics[width=0.9\textwidth]{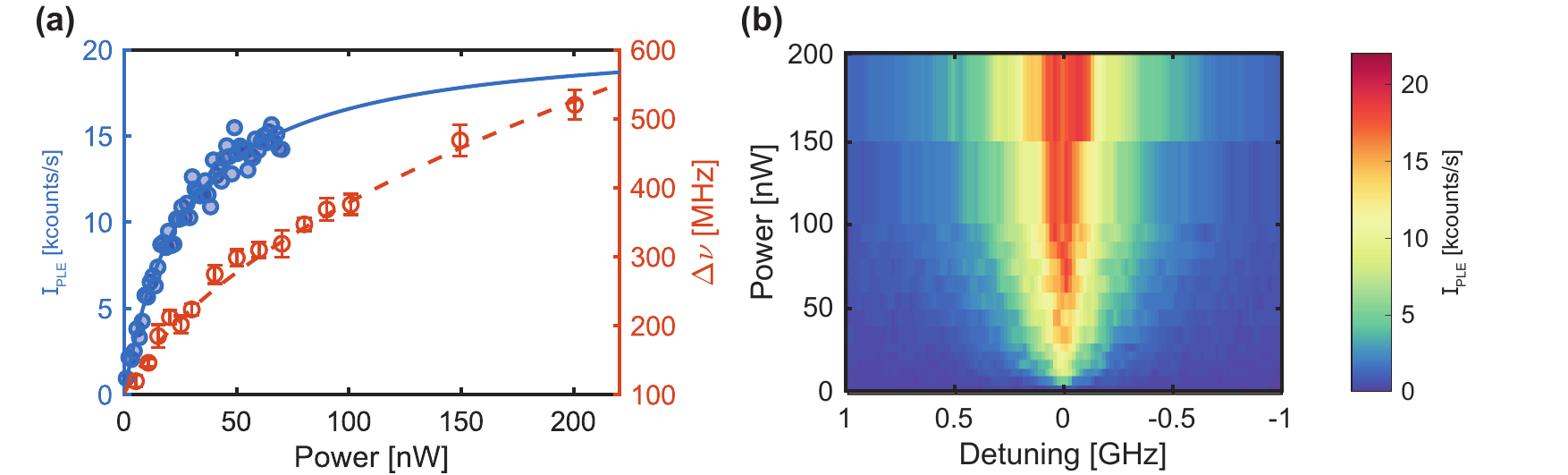}
    \caption{Powerbroadening of an exemplary SiV$^-$. (a) shows the saturation curve measured under resonant excitation of a single SiV$^-$ in PR (blue scatter) with the fitting (blue line) superposed with the linewidth of the emitter measured with increasing excitation powers fitted with the power dependent linewidth $\Delta\nu(P) = \Delta\nu_0\sqrt{1+P/P_{sat}}$, giving the linewidth at zero power, $\Delta\nu_0 = 100.3~$MHz. (b) shows the averaged PLE measurements of this PR with different excitation powers from $1~$nW to $200~$nW.}
 \label{SI_Fig:LWPBEvolution}
\end{figure}

\bibliography{SI/Manuscript1_SiVsInPillars} 